\documentclass[twocolumn,aps,showpacs,nofootinbib,prd]{revtex4}
\usepackage{graphics}
\newcommand{\be}{\begin{eqnarray}}
\newcommand{\ee}{\end{eqnarray}}
\newcommand{\nn}{\nonumber}
\newcommand{\slsh}[1]{{\not \! #1}}
\begin{document}
\title{$\bar{\psi} \psi$-condensate in constant magnetic fields}
\author{M. de J. Anguiano-Galicia, A. Bashir  and A. Raya} 
\affiliation{Instituto de F{\'\i}sica y Matem\'aticas,
Universidad Michoacana de San Nicol\'as de Hidalgo, Apartado Postal
2-82, Morelia, Michoac\'an 58040, M\'exico}

\begin{abstract}

 We solve Dirac equation in the presence of a constant magnetic field
in (3+1)- and (2+1)-dimensions. Quantizing the fermion field, we calculate
$\bar{\psi} \psi$-condensate  from first principles for parity conserving
and violating Lagrangians for arbitrary field strength. We make 
comparison with the results already known in the literature for some
particular cases and point out the relevance of our work for possible 
physical applications.

\end{abstract}

\pacs{11.30.Rd, 11.30.Qc, 12.20.Ds}

\date{\today}

\maketitle


\section{Introduction}

Quantum electrodynamics (QED) exhibits special 
features in the presence of external magnetic fields. For example,
fermions acquire masses dynamically  in the presence of a constant magnetic field of arbitrary strenght for any value of the electromagnetic coupling. This infrared phenomenon, established as universal in (3+1)- and (2+1)- dimensions, is dominated by the
lowest Landau level (LLL) dynamics of electrons, and has 
been dubbed as 
\emph{magnetic catalysis}~\cite{catalysis_1,catalysis_2,catalysis_3,catalysis_4}.
        QED4 in magnetic fields has relevance in the early universe or 
astrophysical environments where physics is likely to be influenced by 
the background magnetic field. On the other hand, 
QED3 has important applications in condensed matter physics, 
{\em e.g.}, in high temperature superconductors~\cite{Supercon1}, 
quantum Hall effect,
and more recently,
Graphene~\cite{graphene}.
In this connection, not only parity conserving but also
parity violating models play a conspicuous role~\cite{FW,Supercon2}. 



In this report, we solve the Dirac equation exactly in the presence of
homogeneous external magnetic field of arbitrary strength for parity 
conserving and 
violating fermionic Lagrangian. 
Expanding out the solutions in their Fourier modes, we calculate
the  $\bar{\psi} \psi$-condensate 
both in (3+1)-dimensions as well as for the generalized case in
(2+1)-dimensions including parity violating mass terms. 
In the LLL approximation for (3+1)-dimensions, our result reduces to that 
of~\cite{GMS} derived through Schwinger proper time method. 
In (2+1)-dimensions, we consider both parity conserving and parity
violating Lagrangians. It is well-known that for the parity conserving case, 
the flavour $U(2)$ 
breaks  down to $U(1) \times U(1)$ spontaneously, even though fermions
do not acquire mass~\cite{GMS,Das}. In other words, they obtain a finite
condensate when $m \rightarrow 0$.
This result is in sharp contrast with
what is observed in (3+1)-dimensions,~\cite{GMS}. Here the condensate
$\simeq m \ {\rm ln}m \rightarrow 0$ as $m \rightarrow 0$.
Therefore, the existence of a finite $\bar{\psi} \psi$-condensate for 
massless bare
fermions has been regarded as a specific (2+1)-dimensional phenomenon.
In this work, we show that for parity non-conserving Lagrangian, 
(2+1)-dimensions do not restrict themselves to this peculiarity.
In the presence of both parity 
conserving and parity violating masses, $m$ and $m_{\tau}$ respectively, 
one can define parity even and parity odd condensates. 
The persistence of these condensates in the corresponding massless limit 
indicates that parity and chiral symmetry get spontaneously violated 
through interactions with the magnetic field. 
We can define convenient linear combinations of these condensates which separate the sectors of different fermion species. We denote them as 
$< \bar{\psi} \psi>_+ $ and $< \bar{\psi} \psi> _-$.  We find that 
 $< \bar{\psi} \psi>_+ \rightarrow
0$ as mass $\rightarrow 0$. Thus, its behaviour is similar to the condensate in (3+1)-dimensions. On the other hand, 
$< \bar{\psi} \psi>_- $ is finite even when mass $\to 0$.

We have organized this brief report as follows~: In section 
II, we solve Dirac equation for all (3+1)- and (2+1)- dimensional cases.
We evaluate corresponding $\bar{\psi} \psi$-condensates in section III. We then present our 
conclusions and discussions.

\section{Dirac equation in magnetic field}

Dirac equation  an external magnetic field is
\begin{eqnarray}
( i\gamma^{\mu}\partial_{\mu} +e\gamma^{\mu}A_{\mu} - m )\psi = 0   \;.
\end{eqnarray}
We first find its solutions in (3+1)-dimensions. As we want to
make a direct connection with (2+1)-dimensions, we adopt the
following representation of $\gamma^\mu$
\be
\gamma^0 =  \left(\begin{array}{cc}
\sigma_3 & 0 \\ 0 & -\sigma_3
\end{array}\right),
\vec{\gamma}= i \left(\begin{array}{cc}
\vec{\sigma} & 0    \\ 0 & -\vec{\sigma}
\end{array}\right),  
\gamma^3 = i \left(\begin{array}{cc}
0 & I \\ I & 0
\end{array}\right),
\label{paridad} 
\ee
where $\vec{\gamma}$ represents matrices $\gamma^1$ and $\gamma^2$. Similarly,
$\vec{\sigma}$ represents Pauli matrices $\sigma_1$ and $\sigma_2$. The
fact that the first three matrices  $\gamma^\mu$, i.e., $\gamma^0$, $\gamma^1$ and
$\gamma^2$ are in the block diagonal form will help make a
comparison between $(3+1)-$ and $(2+1)$-dimensional cases.
We choose the magnetic field in the $z$ direction. Working in the
Landau gauge, we select $A_{\mu} =(0,0,xB,0)$ ($A_y=A^2=-A_{2}=-xB$). 
We can write the positive energy solutions as
\be
\psi^1_{P}&=&N_{n}e^{-i(\vert E_{n} \vert t - py) }
\left(\begin{array}{cccc}
(\vert E_{n}\vert + m)I(n,p,x)\\ \\ -\sqrt{2neB} I(n\!-\!1,p,x) \\ \\ -ip_{z} I(n,p,x) \\ \\ 0
\end{array}\right) \nn \\ \nn \\ \nn \\
\psi^2_{P}&=&N_{n}e^{ -i(\vert E_{n} \vert t - py) }
\left(\begin{array}{cccc}
0 \\ \\ -ip_{z} I(n-1,p,x)\\ \\-\sqrt{2neB} I(n,p,x)\\ \\  (\vert E_{n} \vert + m)I(n\!-\!1,p,x)
\end{array}\right)  \label{eqnpIII}
\ee
\noindent
and the negative energy solutions as
$$
\psi^1_{N}=N_{n}e^{ i(\vert E_{n}\vert t - py) }
\left(\begin{array}{cccc}
 \sqrt{2neB} I(n,-p,x) \\ \\   (\vert E_{n}\vert + m)I(n\!-\!1,-p,x) \\ \\ 0  \\ \\  ip_{z} I(n-1,-p,x)
\end{array}\right)
$$
\be
 \psi^2_{N}=N_{n}e^{i(\vert E_{n} \vert t - py) }
\left(\begin{array}{cccc}
 ip_{z} I(n,-p,x)    \\  \\ 0  \\ \\  (\vert E_{n}\vert + m)I(n,-p,x)   \\ \\   \sqrt{2neB} I(n\!-\!1,-p,x)
\end{array}\right) \label{eqnegIII},
\ee
where
\be
\vert E_n \vert &=& \sqrt{2eBn + m^{2} + p_{z}^2}  \nn \\
N_n &=&\frac{1}{\sqrt{ 2 \vert E_n\vert ( \vert E_n \vert + m  )} }
\nn \\
I(n,p,x) &=& \left(\frac{eB}{\pi}\right)^{1/4} \frac{1}{\sqrt{
{2}^{n}n! }}
H_n\left(\sqrt{eB}\left(x - \frac{p}{eB}\right) \right)  \nn \\
&& \times \; e^{-\frac{eB}{2}\left(x - \frac{p}{eB}\right)^{2} } \;,
\ee
and $I(n=-1,p,x)=0$. Moreover $H_n(x)$ are Hermite polynomials, $p$
is the fermion momentum in the $y$-direction, $n$ labels the Landau
levels and $N_n$ is the normalization factor. We can readily make the
following observations~:
\begin{itemize}

\item In the lowest Landau level (LLL), $\psi^1_{N}=\psi^2_P=0$ trivially. 
Thus LLL is non-degenerate whereas all other levels are degenerate 
in energy. 
This fact is independent of gauge and representation of 
$\gamma^\mu$~\cite{kaushik}.

\item Due to the appropriate choice of representation for the $\gamma$ 
matrices, we do not need to solve the equation for (2+1)-dimensions.
Setting $p_z=0$, we immediately get two decoupled set of solutions
corresponding to two species of fermions in a plane.

\end{itemize}

We now turn to parity violating Dirac equation.
In 4-dimensional representation of planar QED, we have at our
disposal two matrices which commute with all three $\gamma$ matrices
entering the usual Dirac equation. We can take them to be $\gamma^3$
as defined before and $\gamma^5=i \gamma^0 \gamma^1 \gamma^2 \gamma^3$.
\noindent
It permits us to define two types of mass terms~:
$$
{\cal L}
= i\bar\psi\left( \slsh{\partial} -ie\slsh{A} \right)\psi  -m\bar\psi \psi -  m_{\tau}\bar\psi\tau \psi \; ,
$$
where
$$
\tau = \frac{1}{2} [\gamma^3, \gamma^5] =
\left(\begin{array}{cc} I & 0\\ 0 & -I
\end{array}\right) \;.
$$
\noindent
The additional mass term ($m_{\tau}$) in the Lagrangian violates
parity though preserves chiral symmetry, whilst the usual term $m$ is invariant under parity transformations, but breaks chiral symmetry. Corresponding Dirac equation
can be written as~:
$$
(i\gamma^{\mu}\partial_{\mu} + e\gamma^{\mu}A_{\mu} - m  - m_{\tau}\tau)\psi = 0 \; .
$$
We can tabulate the solutions for species $A$ and $B$ of
the fermions as follows~:
\be
\psi^{A}_{P} & = & N_n^{A}  e^{-i(\vert E_n^{A}\vert t - py)}
\left(\begin{array}{cccc}
(\vert E_n^{A} \vert +m_{+}) I(n,p,x) \\ \\ -\sqrt{2eBn} I(n\!-\!1,p,x)\\ \\ 0\\ \\ 0
\end{array}\right) \nn \\ \nn \\ 
\psi^{A}_{N} & = & N_n^{A}  e^{i(\vert E_n^{A} \vert t - py)}
\left(\begin{array}{cccc}
\sqrt{2eBn} I(n,-p,x) \\ \\ ( \vert E_n^{A} \vert + m_{+} ) I(n\!-\!1,-p,x) \\ \\ 0\\ \\ 0
\end{array}\right) \nn \\ \nn \\ 
\psi^{B}_{P} & = & N_n^{B}  e^{-i(\vert E_n^{B} \vert t - py)}
\left(\begin{array}{cccc}
0  \\ \\ 0\\ \\ -\sqrt{2eBn} I(n,p,x)\\  \\ (\vert E_n^{B}\vert + m_{-} ) I(n\!-\!1,p,x)
\end{array}\right) \nn \\ \nn \\ 
\psi^{B}_{N} & = & N_n^{B}  e^{i(\vert E_n^{B} \vert t - py)}
\left(\begin{array}{cccc}
0 \\ \\ 0\\ (\vert{E_n^{B}}\vert + m_{-} ) I(n,-p,x)\\ \\ \sqrt{2eBn} I(n\!-\!1,-p,x)
\end{array}\right) ,
\ee
where
\be
m_\pm &=& m \pm m_\tau, \nn\\
\vert E_n^{A} \vert &=& \sqrt{2eBn + m_{+}^{2} } \;,
N_n^{A} =\frac{1}{\sqrt{ 2 \vert E_n ^{A}\vert ( \vert E_n^{A} \vert +
m _{+}}) }\nn \\ \nn \\ \nn
\vert E_n^{B} \vert &=& \sqrt{2eBn + m_{-}^{2}} \;,
N_n^{B} =\frac{1}{\sqrt{ 2 \vert E_n ^{B}\vert (  \vert E_n^{B} \vert
+m_{-}) } } .
\ee
Note that owing to the parity violation, the fermion spinning
anti-clockwise with respect to its direction of motion has mass
$m_+$, whereas, the fermion spinning clockwise has mass $m_-$.
This difference of masses leads us to new features for the
$\bar{\psi} \psi$-condensate. Furthermore, when $m_\tau=0$, we have $m_\pm=m$, 
which implies $\vert E_n^A\vert=\vert E_n^B\vert$ and $N_n^A=N_n^B$, 
in such a fashion that the above solutions reduce to Eqs.  (\ref{eqnpIII}) 
and (\ref{eqnegIII}) in the limit $p_z=0$, provided we identify  
$A\leftrightarrow 1$ and $B\leftrightarrow 2$.

\section{$\bar{\psi} \psi$-Condensate}

     We first calculate the $\bar{\psi} \psi$-condensate in 3+1 dimensions
by expanding out the fermion field in its Fourier decomposition in
terms of the complete basis of solutions provided by $\{\psi^i_{P,N}\}$~:
\be
\psi(\vec{x},t) &=&
\sum_{n}\sum_{i}^{'}\int \frac{dp}{\sqrt{2\pi}}\int
\frac{dp_z}{\sqrt{2\pi}} \; \times  \nn \\
&& \hspace{0.7cm} \left(a_i(n,p,p_z)\psi^i_{P} + b_i^{\dagger}(n,p,p_z)\psi^i_{N}\right)
, \label{expansion}
\ee
where $i=1,2$ except for the LLL. This is what the primed notation
denotes. The sum over $n$ runs over all Landau levels.
$p$ and $p_z$ are continuous momentum variables. The $a_i$ and $b_i$
are the particle and anti-particle destruction operators respectively
obeying the anti-commutation relations~:
\be
\{a_i(n,p,p_z),a_j^\dagger(n',p',z')\}&=& \{b_i(n,p,p_z),b_j^\dagger(n',p',z')\}\nn\\
& =& \delta_{ij} \delta_{nn'} \delta(p-p') \delta(p_z-p_z'). \nn \\ \nn
\ee
Evaluation of the condensate thus leads to:
\be
\langle \bar{\psi} \psi \rangle &=& - \frac{m}{(2\pi)^2} \sum_n 
\int_{-\infty}^{\infty} dp \left[ I^{2}(n-1,p,x) + I^{2}(n,p,x) \right]\nn\\
&\times&
\int_{-\infty}^{\infty} dp_{z}\frac{1}{\sqrt{ m^2 +p_{z}^2 + 2neB }} \; .\label{cond-4}
\ee
\\ \nn \\
The integral and the sum over Landau levels in Eq.~(\ref{cond-4}) are divergent and we need to
introduce cut-offs for both. Moreover, we separate the LLL contribution from the rest 
of the Landau levels and employ the normalization conditions for the Hermite polynomials to 
arrive at
\be
\langle \bar{\psi} \psi \rangle &=& - \frac{meB}{2 \pi^2}  {\rm ln} 
\left[ \frac{1+\sqrt{1+x_0}}{\sqrt{x_0}} \right]  \nn \\
&& - \frac{meB}{\pi^2} \sum_{n=1}^{n_{\rm max}}  {\rm ln} \hspace{-1mm}
\left[ \frac{1+\sqrt{1+x_n}}{\sqrt{x_n}} \right], \label{sum-4D}
\ee
where $x_n = (m^2+ 2 n e B)/\Lambda^2$. 
Through the method of Schwinger proper time, this condensate in 
terms of an integral over the proper time was obtained in~\cite{GMS2}. As expected, 
$\langle \bar{\psi} \psi \rangle \rightarrow 0 $ as
$m \rightarrow 0$. Although the sum over the Landau levels diverges, contribution of individual
levels converges for growing $n$ as indicated by the fact that
$$
 n \rightarrow  \infty \qquad  \Rightarrow \qquad \ln\left( 
\frac{1 + \sqrt{1 + x_n} }{\sqrt{x_n}}\right)
\rightarrow 0 \; .
$$
One can see that in the limit of $m$ very small, the leading contribution to the condensate comes from $n=0$~:
\be
\langle \bar{\psi} \psi \rangle = - \frac{m eB}{4 \pi^2}  {\rm ln} \frac{\Lambda^2}{m^2}  \; .
\ee
Thus in the LLL approximation, we
recuperate the expression obtained in~\cite{GMS}. 

In (2+1)-dimensions, two types of mass terms ($m$ and $m_{\tau}$) are
connected to two types of condensates $\langle \bar{\psi} \psi
\rangle$ and $\langle \bar{\psi} \tau \psi \rangle $. We again
resort to the second quantized representation of $\psi$ field.
\be
\psi(\vec{x},t) =
\sum_{n}\sum_{i}^{'}  \hspace{-1mm}   \int \hspace{-1mm} 
\frac{dp}{\sqrt{2\pi}} \left(a_i(n,p)\psi^i_{P} + b_i^{\dagger}(n,p)\psi^i_{N}\right),
\ee
where $i=A,B$. This yields~:
\be
\langle  \bar{\psi} \psi  \rangle & =&  -\frac{eB}{2\pi}\frac{m_{-}}{\vert m_{-} \vert}
 -\frac{eB}{2\pi}\sum_{n=1}^{\infty}\left( \frac{m_{+}}
{\vert E_{n}^{A} \vert} + \frac{m_{-}}
{\vert E_{n}^{B} \vert}\right)  \nn \\
\langle \bar{\psi}\tau \psi  \rangle &=&  \frac{eB}{2\pi}\frac{m_{-}}{\vert m_{-} \vert}
 -\frac{eB}{2\pi}\sum_{n=1}^{\infty}\left( \frac{m_{+}}
{\vert E_{n}^{A} \vert} - \frac{m_{-}}
{\vert E_{n}^{B} \vert}\right) \; .\label{PEPOcond} 
\ee
Note that  when $m=0$, $\langle \bar{\psi} \psi  \rangle\ne 0$ and thus 
chiral symmetry is broken by the interections of electrons with the external magnetic field. Also,  
in the case $m_{\tau}=0$, $\langle \bar{\psi}\tau \psi  \rangle$ is non vanishing.  This reflects that 
in QED3, these interactions  lead to the violation of parity as well. Furthermore, in the latter case, 
the parity conserving condensate reduces to the result of Das~\cite{Das}~:
\be
\langle  \bar{\psi} \psi  \rangle & =&  -\frac{eB}{2\pi}\frac{m}{\vert m \vert}
 -\frac{eBm}{\pi}\sum_{n=1}^{\infty}  \frac{1}
{\vert E_{n} \vert} \label{PC-3D} \;.
\ee
At this point, it is convenient to notice that the projectors $(1\pm\tau)/2$ allow us to entirely separate 
the $A$ and $B$ sectors of the theory~\cite{kondo}. This, in turn, permits us to write the following linear 
combinations of the condensates:
\be
\langle  \bar{\psi} \psi  \rangle_+ &=& \langle  \bar{\psi} \psi
\rangle 
+ \langle \bar{\psi}\tau \psi  \rangle
\nn \\
\langle  \bar{\psi} \psi  \rangle_- &=& \langle  \bar{\psi} \psi
\rangle - 
\langle \bar{\psi}\tau \psi  \rangle
\nn \; .
\ee
Using the previous relations,
\be
\langle  \bar{\psi} \psi  \rangle_+ &=&
-\frac{eB}{\pi}\sum_{n=1}^{\infty}\frac{m_{+}}{|E_{n}^{A}|}
 \nn \\
\langle  \bar{\psi} \psi  \rangle_- &=&  
-\frac{eB}{\pi}\frac{m_{-}}{|m_{-}|}  -\frac{eB}{\pi}
\sum_{n=1}^{\infty}\frac{m_{-}}{|E_{n}^{B}|}
 \; . \label{PVcond-3D}
\ee
Consequently, $\langle  \bar{\psi} \psi  \rangle_+ \rightarrow 0$ as
$m_+ \rightarrow 0$, a feature usual to condensates in (3+1)-dimensions. However, $\langle  
\bar{\psi} \psi  \rangle_-$ retains a
finite value as $m_- \rightarrow 0$. It is
\be
\langle  \bar{\psi} \psi  \rangle_- =  -\frac{eB}{\pi}{\rm sgn}(m_-) 
\label{zeromass} \;.
\ee 
This compares to the fact that massless theory exhibits a current of abnormal parity
\be
\langle J^\mu(x) \rangle = {\rm sgn}(m) \frac{e^2}{4\pi} \ ^* F^\mu(x),
\ee
where $^* F^\mu(x)=(1/2) \epsilon^{\mu\alpha\beta}F_{\alpha\beta}$ is
the dual stress tensor,~\cite{cap}. 

An important point to notice is the indeterminacy of the value of the condensates (\ref{PEPOcond}) and (\ref{zeromass}) when $m=m_\tau$. A similar feature was found in Ref.~\cite{zhuko}. In this work, lack of differentiability of the effective potential of a model of QED3 with parity conserving and violating Yukawa mass terms was found along the line where both mases are equal. 

Finally, expanding the condensates in the strong field limit, we obtain
\be
\langle  \bar{\psi} \psi  \rangle_- &=& 
-\frac{eB}{\pi}\frac{m_{-}}{|m_{-}|} +
 \sqrt{ \frac{eB}{2} } \;   \frac{m_-}{2 \pi}
\; \zeta(3/2) +   {\cal O} \left( \frac{m_-^2}{eB} \right)  \nn \\
\langle  \bar{\psi} \psi  \rangle_+ &=&  \sqrt{ \frac{eB}{2} } \;   \frac{m_+}{2 \pi}
\; \zeta(3/2) +   {\cal O} \left( \frac{m_+^2}{eB} \right) \;.  \label{strongB-3D}
\ee

The main results of the report are Eqs.~(\ref{sum-4D}, \ref{PEPOcond}
\ref{PVcond-3D}, 
\ref{zeromass}, \ref{strongB-3D}). 
These equations relate $\bar{\psi} \psi$-condensate with the 
fermion mass and the magnetic field in (3+1)-dimensions and (2+1)-dimensions for 
parity conserving and violating Lagrangians. 
The result in (3+1)-dimensions can have application in the physics of the early universe where the possible existence
of magnetic/hypermagnetic fields has been shown to modify the behaviour of
the electroweak phase transition, see for example~\cite{Alejandro2} and references
therein. On the other hand, our findings for (2+1)-dimensions are of relevance in
 parity conserving and violating effective theories describing high 
temperature superconductivity and the quantum Hall effect, where a magnetic field is
an important ingredient of the dynamics. In particular, these could be of direct
importance in the
study of graphene,
where in general, a kaleidoscopic variety of mass terms and 
order parameters can be defined which might manifest themselves in the 
experiment~\cite{graphene1}. 
Work in this direction is in progress.
Moreover, carrying out this work in the presence of
a heat bath is likely to be of even greater interest in the
study of the above mentioned physical scenarios. 
All this is for future.

\noindent
{\bf Acknowledgments}


We acknowledge R. Delbourgo and V. Gusynin for valuable comments.  AB and 
AR acknowledge SNI as well as COECyT, CIC and CONACyT grants under projects CB070218-4, 
CB0702152-4, 4.12, 4.22 and 46614-I.

\hsize=16.5cm

\vfil\eject
\vskip 1cm

\baselineskip=8mm
\begin{thebibliography}{999}

\bibitem{catalysis_1} V.P. Gusynin, V.A. Miransky and I.A. Shovkovy, Phys.
Lett. {\bf B349} 477 (1995); Phys. Rev. {\bf D52} 4747 (1995); Nucl.
Phys. {\bf B462} 249 (1996); Nucl. Phys. {\bf B563} 361 (1999).
%
%
\bibitem{catalysis_2} D.-S. Lee, C.N. Leung and Y.J. Ng, Phys. Rev. 
{\bf D55} 6504 (1997).
%
%
\bibitem{catalysis_3} D.K. Hong, Phys. Rev. {\bf D57} 3759 (1998).
%
%
\bibitem{catalysis_4} E.J. Ferrer and V. de la Incera, Phys. Lett. 
{\bf B481} 287 (2000).
%
%
%
%
%
%
%
%
%
%
%
%
%
\bibitem{Supercon1}  M. Franz and Z. Tesanovic, Phys. Rev. Lett. 
{\bf 87} 257003 (2001);
 M. Franz, Z. Tesanovic and O. Vafek, Phys. Rev. {\bf B66}
054535 (2002); O. Vafek, Z. Tesanovic and M. Franz, Phys. Rev. Lett. 
{\bf 89} 157003 (2002); I.F. Herbut, Phys. Rev. {\bf B66} 094504 (2002);
 I.F. Herbut, Phys. Rev. Lett. {\bf 88} 047006 (2002).
%
\bibitem{graphene} K.S. Novoselov, A.K. Geim, S.V. Morozov, D. Jiang M.I. Katnelson, 
I.V. Grigorieva, S.V. Dubonos and A.A. Firsov, Nature {\bf 438} 197 (2005); Y. Zhang, 
Y.-W. Tan, H.L. Stormer and P. Kim, Nature {\bf 438} 201 (2005); V.P. Gusynin 
and S.G. Sharapov, Phys. Rev. Lett. {\bf 95} 146801 (2005); V.P. Gusynin and S.G. 
Sharapov, Phys. Rev. {\bf B73} 245411 (2006).
%
\bibitem{FW} E. Fradkin, \emph{``Field Theories of Condensed Matter Systems''}
(Addison-Wesley Publishing Company, (1991); F. Wilczek, \emph{``Fractional
Statistics and Anyon Superconductivity''}, World Scientific (1998); A. Khare, \emph{``
Fractional Statistics and Quantum Theory''}, World Scientific (2005).

\bibitem{Supercon2} N. Dorey and N. E. Mavromatos,  Nucl. Phys. {\bf
B386} 614 (1992);  G. Triantaphyllou, Phys. Rev. {\bf D58} 065006 (1998).

\bibitem{GMS} V. P. Gusynin, V. A. Miransky and I. A. Shovkovy,
Phys. Rev. {\bf D52} 4718 (1995).

\bibitem{Das} A. Das, \emph{``Finite Temperature Field Theory''} World Scientific Publishing Company, (1997).


\bibitem{kondo} K.-I. Kondo and P. Maris, Phys. Rev. {\bf D52} 1212 (1995);
 K.-I. Kondo and P. Maris, Phys. Rev. Lett. {\bf 74} 
18 (1995).

\bibitem{kaushik} K. Bhattacharya, \emph{``Solution of the Dirac equation in presence of an uniform magnetic field''}, e-Print: arXiv:0705.4275 [hep-th].

\bibitem{GMS2} V. P. Gusynin, V. A. Miransky and I. A. Shovkovy,
Phys. Lett. {\bf B349} 477 (1995).


\bibitem{cap} L. Alvarez-Gaum\'e and E. Witten, Nucl. Phys. {\bf B234} 269 (1983); A.J. Niemi and G.W. Semenoff, Phys. Rev. Lett. {\bf 51} 2077 (1983); A.N. Redlich, Phys. Rev Lett. {\bf 52} 18 (1984); P. Cea, Phys. Rev. {\bf D55} 7985 (1997).

\bibitem{zhuko} V. Ch. Zhukovsky, K.G. Klimenko and V.V. Khudyakov, Theor. Math. Phys. {\bf 124} 1132 (2000); Teor. Mat. Fiz. {\bf 124} 323, (2000). 

\bibitem{Alejandro2} A. S\'anchez, A. Ayala and G. Piccinelli, Phys. Rev. {\bf D75} 043004 (2007).

\bibitem{graphene1} V.P. Gusynin, S.G. Sharapov and J.P. Carbotte, {\em ``AC Conductivity of Graphene: From Tight Binding Model to 
2+1-Dimensional Quantum Electrodynamics''}, e-Print: arXiv:0706.3016 [cond-mat.mes-hall]; I.F. Herbut, Phys. Rev. {\bf B 76}, 085432 (2007).
\end{thebibliography}
\end{document}